\documentclass[a4paper]{jpconf}
\usepackage{graphicx}
\usepackage{slashed}
\usepackage{amsmath}
\usepackage{amsfonts}
\usepackage{amssymb}
\usepackage{wrapfig}
\usepackage{hyperref}
\usepackage{color}
\usepackage{subfig}

\begin{document}

\begin{flushright}
WITS-MITP-024, HRI-RECAPP-2016-002
\end{flushright}

\title{The impact of additional scalar bosons at the LHC}

\author{Mukesh Kumar$^{a, 1}$,  
  Stefan von Buddenbrock$^{b, 2}$, Nabarun Chakrabarty$^{c, 3}$,
  Alan S. Cornell$^{a, 4}$, Deepak Kar$^{b,5}$, Tanumoy Mandal$^{d,6}$, Bruce Mellado$^{b,7}$, 
  Biswarup Mukhopadhyaya$^{c,8}$ and Robert G. Reed$^{b,9}$}
  
\address{$^{a}$ National Institute for Theoretical Physics; School of Physics and Mandelstam Institute for Theoretical Physics, University of the Witwatersrand, Johannesburg, Wits 2050, South Africa}
\address{$^{b}$ School of Physics, University of the Witwatersrand, Johannesburg, Wits 2050, South Africa}
\address{$^{c}$ Regional Centre for Accelerator-based Particle Physics, Harish-Chandra Research Institute, Chhatnag Road, Jhusi, Allahabad - 211 019, India.}
\address{$^{d}$ Department of Physics and Astronomy, Uppsala University, Box 516, SE-751 20 Uppsala, Sweden.}
\ead{$^{1}$mukesh.kumar@cern.ch, 
$^{2}$stef.von.b@cern.ch, 
$^{3}$nabarunc@hri.res.in,
$^{4}$alan.cornell@wits.ac.za,
$^{5}$deepak.kar@cern.ch,
$^{6}$tanumoy.mandal@physics.uu.se,
$^{7}$bruce.mellado@wits.ac.za,
$^{8}$biswarup@hri.res.in, 
$^{9}$robert.reed@cern.ch}

\begin{abstract}
In this study we consider an effective model by introducing two hypothetical real scalars, 
$H$ and $\chi$ - a dark matter candidate, where the masses of these scalars are $2 m_h < m_H < 2 m_t$ 
and $m_\chi \approx m_h/2$ with $m_h$ and $m_t$ being the Standard Model Higgs boson and 
top quark masses, respectively. A distortion in the transverse momentum distributions of $h$ in the
intermediate region of the spectrum through the 
processes $p p \to H \to h\chi\chi$ could be observed in this model.
An additional scalar, $S$, has been postulated to explain large $H \to h\chi\chi$
branching ratios, assuming $m_h \lesssim m_S \lesssim m_H-m_h$ and $m_S > 2 m_\chi$.
Furthermore, a scenario of a two Higgs doublet model (2HDM) is introduced and a detailed proposal at the 
present energies of the Large Hadron Collider to study the extra CP-even ($h, H$), CP-odd ($A$) and
charged ($H^\pm$) scalars has been pursued. With possible phenomenological implications,
production and decay modes for these scalars are discussed. 
Based on the mass spectrum of $H, A$ and $H^\pm$, the production of multi-leptons and
$Z$+jets+missing-energy events are predicted.
A specific, Type-II 2HDM model is discussed in detail. 
\end{abstract}

\section{Introduction}
\label{intro}

With the discovery of the scalar Higgs boson, $h$, the particle spectrum of the Standard Model (SM) is now complete. 
But there are numerous theoretical motivations to expect that the SM electroweak symmetry  breaking
(EWSB) mechanism is not the complete story, and the scalar spectrum responsible for it may be 
richer than one possessing only one neutral scalar - the Higgs boson. From this expectation,
searches for new scalars, neutral or charged, are continuously being carried out both by the ATLAS and
 CMS collaborations in various channels. No confirmed hints along these directions has been found so far in 
any of these searches.

The nature of these additional particles in different theories are based on its charge, spin, mass and interactions 
with the SM particles. Any beyond the SM (BSM) scenario can be seen in various observables associated with the 
decay modes of these new particles. For example, any enhancement in the fiducial cross sections for the relevant 
processes could cause excesses in transverse momentum, $p_T$, spectrum in the intermediate regions of the decayed 
particles, in (pseudo)rapidity distributions or in some angular observables. Signatures of a massive object with an 
appropriate choice of its decay width can also be observed as an excesses in the invariant mass distribution around 
the mass of that object. These excesses can be understood with the help of appropriate theoretical models. 

It is also noted that, if any new physics is present, one would expect that it will possibly pop-up, not only in one set
of data, but in different (theoretically) correlated searches or channels with different significances. Therefore, while 
interpreting the excesses or anomalies, it is important to scrutinise different data sets simultaneously, which we think
might be correlated in terms of simple model assumptions.
  
In these proceedings, we shall discuss a formalism to address some of these subjects by extending the SM with another 
Higgs-doublet in the theory and with an additional scalar or fermions to make the modelling appropriate and obtain the 
correct kinematics. 
Section~\ref{theory} will briefly deal with the theoretical model by extending the SM with one Higgs-doublet. 
The phenomenological aspects, with proposed formalism, are discussed in section~\ref{pheno}, and summary follows 
in section~\ref{conc}. 
   
\section{Theory - adding one more Higgs-doublet}
\label{theory}
We start by quoting one of the classic texts in Higgs-boson physics~\cite{gunion} -

{\it{``In this book we will repeatedly emphasise that many people who have studied Higgs physics do not 
expect a single, neutral boson of some definite mass to occur as the only direct manifestation of 
electroweak symmetry  breaking (EWSB)."}}

As we all know the ``minimal Higgs'' in the $SU(2)_L \times U(1)_Y$ electroweak theory Higgs sector is
comprised of only one complex Higgs doublet, and we have only one physical neutral Higgs scalar present 
in the SM spectrum, where its mass is a free parameter, not fixed by theory. In the current
scenario, where this Higgs boson is almost confirmed to be $m_h \approx$ 125 GeV, it is prudent to 
explore the implications of more complicated Higgs models in the context of extended theories.
In any such theories there are basically two major constraints which need to be satisfied:
\begin{itemize} 
\item [(a)] First constraint comes from the experimental fact that $\rho = m_W^2 / (m_Z^2 \cos^2\theta_W)$ 
should be very close to 1. The general formula is
\begin{equation}
\rho \equiv \frac{m_W^2}{m_Z^2 \cos^2\theta_W} = 
\frac{\sum_{T,Y} \left[ 4\,T\left(T+1\right)-Y^2 \right] \left| V_{T,Y}  \right|^2 c_{T,Y}}{\sum_{T,Y} 2\,Y^2 \left| V_{T,Y}  \right|^2},
\end{equation}
where $\langle \Phi \left(T,Y\right)\rangle = V_{T,Y}$ defines the vacuum expectation values of each neutral
Higgs field $\Phi$, with $T$ and $Y$ specifying the total $SU(2)_L$ isospin and hypercharge of the Higgs 
representation to which it belongs. Here the notation, $c_{T,Y} = 1$ is for complex representation of ($T, Y$) and
$c_{T,Y} = 1/2$ should be for real representation of ($T, Y = 0$). 
\item[(b)]  Second major theoretical constraint on the Higgs sector comes from severe limits on the existence 
of flavour-changing neutral currents 
(FCNC's), which is absent in the minimal Higgs model at tree level and which must also be true in any extended models. 
\end{itemize}
 Keeping these issues in mind, let's extend by one more Higgs doublet and call it a two Higgs doublet model (2HDM).
 To date there are many versions or types of 2HDMs~\cite{Branco:2011iw} and the key features can be summarised
 as follows:
 \begin{itemize}
 \item [(1)] As an extension of the SM, this model adds new phenomena (e.g. the physical charged Higgs bosons).
 \item [(2)] Since it is a minimal extension, it adds only a few new arbitrary parameters.
 \item [(3)] It satisfies theoretical constraints of $\rho \approx 1$ and the absence of potentially dangerous tree-level
 FCNC,\footnote{
 If the Higgs-fermion couplings are appropriately chosen, however, based on current FCNC requirements 
 in some specific scenarios, this also allowed at tree level~\cite{Branco:2011iw}.} by imposing some particular symmetries.
 \end{itemize}
 Considering two complex $SU(2)_L$ doublet scalar fields $\Phi_1$ and $\Phi_2$, the most general remormalizable
 scalar potential may be written as:
 \begin{align}
 V(\Phi_1, \Phi_2) = &\,\, m_1^2 \Phi_1^\dag \Phi_1 + m_2^2 \Phi_2^\dag \Phi_2 
 - m_{12}^2 \left(\Phi_1^\dag \Phi_2 + \text{h.c.} \right) + \frac{1}{2} \lambda_1 \left(\Phi_1^\dag \Phi_1\right)^2
 + \frac{1}{2} \lambda_2 \left(\Phi_2^\dag \Phi_2\right)^2  \notag\\
 & + \lambda_3 \left(\Phi_1^\dag \Phi_1\right) \left(\Phi_2^\dag \Phi_2\right)
 + \lambda_4 \left| \Phi_1^\dag \Phi_2\right |^2 + \frac{1}{2} \lambda_5 \left[  \left( \Phi_1^\dag \Phi_2 \right)^2 + \text{h.c.} \right ] \notag \\
 & + \left[ \left[ \lambda_6 \left(\Phi_1^\dag \Phi_1\right) + \lambda_7 \left(\Phi_2^\dag \Phi_2\right)\right]
 \Phi_1^\dag \Phi_2 + \text{h.c.}\right].
 \label{pot2hdm}
 \end{align}
 After spontaneous breaking of the electroweak symmetry, five physical Higgs
 particles are left in the spectrum, one charged Higgs pair, $H^\pm$, one CP-odd scalar, $A$, and two CP-even states,
 $H$ (heaviest) and $h$ (lightest) given as:
 \begin{align}
 &H^\pm = \sin\beta \,\phi_1^\pm + \cos\beta \,\phi_2^\pm, \\
 &A = \sin\beta \,\text{Im} \,\phi_1^0 + \cos\beta \,\text{Im}\, \phi_2^0, \\
&H = \cos\alpha \left(\text{Re}(\phi_1^0)-v_1\right) + \sin\alpha \left(\text{Re}(\phi_2^0)-v_2\right), \\
&h = -\sin\alpha \left(\text{Re}(\phi_1^0)-v_1\right) + \cos\alpha \left(\text{Re}(\phi_2^0)-v_2\right).
 \end{align}
 Here $\phi_i^+$ and $\phi_i^0$ denote the $T_3 = 1/2$ and $T_3 = -1/2$ components of the $i^{th}$ doublet for
 $i = 1, 2$. The angle $\alpha$ diagonalises the CP-even Higgs squared-mass matrix and 
 $\beta$ diagonalises both the CP-odd and charged Higgs sectors with $\tan\beta = v_2/v_1$, where
 $\langle \phi_i^0 \rangle= v_i$ for $i = 1, 2$ and $v_1^2 + v_2^2 \approx (246\,\text{GeV})^2$. This is a brief  
 summary of how 2HDMs are constructed and further based on different choices of symmetries, couplings to quarks 
 and leptons etc. different models can be built. Models which lead to natural flavour conservation can be named as Type-I, Type-II, 
 Lepton-specific or Flipped 2HDMs, as detailed in Ref.\cite{Branco:2011iw}. In our studies we used a Type-II
 2HDM for reference.

\section{Phenomenology and formalism}
\label{pheno}
In the previous section we discussed an extension of the Higgs boson sector, where apart from the SM Higgs 
boson we have extra particle spectrum of phenomenological interest. Here we describe an effective Lagrangian 
formalism and then follow a discussion on a specific scenario of a 2HDM, with prospects to correlate these studies 
for phenomenological purposes. 

In all numerical analyses the codes are written in {\texttt {Mathematica}}, {\texttt {C$^{++}$}}, {\texttt {Python}}, 
models are generated with {\texttt {FeynRules}} which are further used for simulation through {\texttt {MadGraph5}}. 
Events are then showered and hadronized using {\texttt {Pythia 8.2}}, plotting and analyses 
followed through {\texttt {Rivet}} and {\texttt {ROOT}} package. 
Dark matter related constraints calculated using {\texttt {microOMEGAs}}. 

\subsection{An effective Lagrangian approach}
\label{effth}
Here we consider an effective Lagrangian approach with the introduction of two hypothetical real scalars 
$H$ and $\chi$ beyond SM particle spectrum, to study phenomenology associated with Higgs boson physics.  
The formalism considered heavy scalar boson which decays into the SM Higgs and $\chi$,
where $\chi$ is considered as a dark matter (DM) candidate - a source of missing energy (MET). 
The required vertices for the studies are:
\begin{align}
\mathcal{V}_{H} &= -\frac{1}{4}~\beta_{g} \kappa_{_{hgg}}^{\text{SM}}~G_{\mu\nu}G^{\mu\nu}H
+\beta_{_V}\kappa_{_{hVV}}^{\text{SM}}~V_{\mu}V^{\mu}H,  \label{vh} \\
\mathcal{V}_{\text{Y}} &= -\frac{1}{\sqrt{2}}~\Big[y_{_{ttH}}\bar{t} t H + y_{_{bbH}} \bar{b} b H\Big],\\ 
\mathcal{V}_{\text{T}} &=-\frac{1}{2}~v\Big[\lambda_{_{Hhh}}Hhh + \lambda_{_{h\chi\chi}}h\chi\chi + \lambda_{_{H\chi\chi}}H\chi\chi\Big], \\
\mathcal{V}_{\text{Q}} &= -\frac{1}{2}\lambda_{_{Hh\chi\chi}}Hh\chi\chi - \frac{1}{4} \lambda_{_{HHhh}}HHhh 
-\frac{1}{4}\lambda_{_{hh\chi\chi}}hh \chi\chi -\frac{1}{4} \lambda_{_{HH\chi\chi}}HH\chi\chi, \label{vq}
\end{align}     
where $\beta_g = y_{ttH}/y_{tth}$ is the scale factor with respect to the SM Yukawa top coupling, $y_{tth}$ and 
used to tune the effective $ggH$ coupling as $\beta_V$ is used for $VVH$ couplings. Major constraints on the 
parameters of the vertices come from the relic density of DM and the DM-nuclei inelastic scattering cross sections, 
which left a narrow choice of mass of the DM candidate $m_\chi \sim m_h/2$  with $\lambda_{h\chi\chi} \sim [0.0006-0.006]$.     
Therefore, the masses of these scalars considered to be $2 m_h < m_H < 2 m_t$ and $m_\chi \approx m_h/2$, 
where $m_t$ is the top quark mass.  
In this study if we consider a process $pp \to H \to h\chi\chi$, a distortion could be predicted in the intermediate 
Higgs $p_T$ spectrum due to decay of heavy $H$ and a recoil affect against a pair of invisible particles, $\chi$.
In order to chose appropriate values of associated couplings one must consider the constraints from all potential
sources and different processes.   
It is to be noted, while considering various processes where the effective $Hh\chi\chi$ coupling plays a vital 
role one may assume this coupling to be large or small in respect of branching ratios for $H \to h \chi\chi$. 

However in this effective field theory approach, we are not sure about the origin of the $Hh\chi\chi$ coupling. 
And in case if it is large, we can assume that this term can be generated due to the participation of a real 
scalar particle, $S$, in the intermediate state through an effective $HhS$ coupling followed by the decay of 
$S\to\chi\chi$. Therefore, the inclusion of $S$ can open up various new possibilities.
In addition to the above studies, if we look over the double Higgs production modes in different decay 
channels like $\gamma\gamma b \bar b$ or $b\bar b b \bar b$ with jets etc., the vertices defined above
(\ref{vh})-(\ref{vq}) will be modified appropriately with $S$ as an intermediate scalar and not a DM 
candidate,\footnote{$S$ is a scalar particle with various decay modes and having all possible 
branchings to other particles, since this is no more longer stable like a DM candidate.
As a result, the symmetry requirements for a gauge invariant set of vertices in the Lagrangian is different.}
with $m_h \lesssim m_S  \lesssim m_H - m_h$ and $m_S > 2m_\chi$.  This invokes other 
possibilities for the mass spectrum of $S$. The processes in such studies include $pp \to hS + X$ and 
in particular $pp \to H\to hS + X$ with respect to $pp \to H \to hh + X$, considering the available
spectrum of $m_S$ and the associated coupling parameters.
There is a possibility to introduce a $HSS$ vertex in the study, which participates further in a $H \to S S$ 
decay channel, as $H \to h h$ modes, with the participation of this vertex in other possible processes too.
Furthermore, all other possible decay modes of $S$, with $S$ into jets are possible.

\subsection{Phenomenology}
\label{pheno2hdm}

Considering these analyses, now we can think of $H$ as a heavy scalar\footnote{It is to be noted that in 
the studies of subsection~\ref{effth} through the effective Lagrangian, 
the scalar $H$ is not necessarily assumed to be a 2HDM heavy scalar.} 
of the 2HDM, and furthermore, our motive
should then be to fit all the parameters like $\tan\beta$, $\alpha$ and the masses of $A$ and $H^\pm$ in this specific
model. But it is also a question as to whether we think of a generalised version of a 2HDM or any particular
type of this model, as described in detail in Ref.~\cite{Branco:2011iw}. On the other hand, we also need to consider
experimental data on searches which will directly or in-directly affect the processes taken into consideration in
this model and the signatures could be:
\begin{itemize}
\item [(1)] The resonance searches gives an excess around a particular narrow mass range (with appropriate decay 
width approximation) 
in the invariant mass spectrum in di-jets or di-boson, $VV$, (where $V = \gamma, Z, W^\pm$) final states, and
that provides a hint for a beyond SM particle. 
The masses for these resonances, $m_\Phi$, (where $\Phi = H, A , H^\pm$ in 2HDMs) 
might be of the order of $2 m_h < m_\Phi < 2 m_t$ 
(which we considered in our previous studies for $m_\Phi = m_H$) or 
beyond this order like $2 m_t \ll m_\Phi < $ $\mathcal{O}$(1 TeV) or  even $m_\Phi \gg \mathcal{O}$(1 TeV). 
\item [(2)] In the 2HDM spectrum, we also have the possibility of charged Higgs ($H^\pm$) that can be produced at 
the LHC and these searches follows considering production cross sections and branching ratios in different decay
channels. 
The prominent modes of decay of $H^{\pm}$ are $H^\pm \to tb, W^\pm h$ when $m_{H^\pm} > m_t$. 
Since, we consider $2 m_h < m_H <  2 m_t$, the decay mode of $H^\pm \to W^\pm H$ could be a prominent 
channel too in case of $m_{H^\pm} \gg m_H$. 
\end{itemize}

In purview of phenomenological aspects for a CP-odd scalar, $A$, in 2HDMs, following salient features could be observed:
\begin{itemize}
 \item [(1)] In 2HDMs masses of $A$ and $H^\pm$ are correlated. 
 So if we wish to have a 2HDM with a particular mass $m_A$, comparably $m_{H^\pm}$ should also be
 considered. And with a known value of $m_H$ ($2 m_h < m_H < 2 m_t$) and $m_h = 125$ GeV one should tune the parameters 
 $\alpha$ and $\beta$ accordingly.
 \item [(2)] In case of a production mode of $A$, through the $ggA$ vertex, there will be a need to have a scale factor 
 $\beta_g^A$ as taken in the case of $H$ production, $\beta_g$, and considering the decay modes $A \to \gamma\gamma$ 
 in particular needs another scale factor $\beta_\gamma^A$. In this respect, for consistency, one needs to control the 
 $H\to \gamma\gamma$ 
 decay rates via another parameter $\beta_\gamma$, since the form factors appearing in the calculation of $gg \to H, A$ and
 $H, A \to \gamma\gamma$ have a different structure and also are dependent on the masses of particles under consideration
 \cite{gunion}. One should also study other possible decay modes $A$ in $W^\pm$ or $Z$ which is available only at
 loop level in 2HDMs, since $W^+W^-A$ and $AZZ$ couplings are absent due to CP conservation issues.
 \item[(3)] This model can predict anomalously large $Z$+jets+MET events, we need to think of the contribution of the
 decay mode of $A \to Z H$, where the process should be $p p \to A \to ZH, H \to h\chi\chi$ ($\chi$ should be considered
 as missing energy, similar to what we considered before in $p p \to H \to h \chi\chi$). To be noted, here we need
 $m_A > m_Z + m_H$.
\item [(4)] In respect of point (3) mentioned above, we also consider different processes with final states with multi-lepton, 
tri-lepton and di-lepton channels 
by same-sign and opposite-sign leptons selection with jets in our phenomenological interests due to the
charged Higgs, $H^\pm$, in 2HDMs.
 \item[(5)] Since Yukawa couplings for top-quarks, $y_{tth}$, are well known, one should adjust the parameters $\alpha$ and
 $\beta$ in such a way so that $y_{At\bar t}$ and $y_{Ht\bar t}$ must follow appropriate branchings into $A\to t \bar t$
 and $H \to t \bar t$. Here we should also notice that, since $y_{tth}$ is close to unity due to large top-quark mass,
 it also adds an insight for the scale of any new physics.     
\end{itemize}

The phenomenology of $H^\pm$ itself is a subject of great detail, since one should consider either $m_{H^\pm} < m_t$ 
or $m_{H^\pm} > m_t$. And due to this fact the decay modes for the studies are dependent on $m_{H^\pm}$ following the
appropriate mixing parameters $\alpha$ and $\beta$ in 2HDMs. We consider the case in which $m_{H^\pm} > m_t$.
The production of $H^\pm$ at the LHC would follow two mechanisms which have sizeable production cross sections 
(1) $2 \to 2$, $p p \to gb(g\bar b) \to tH^- (\bar tH^+)$ and 
(2) $2 \to 3$, $p p \to gg/qq^\prime \to t H^- \bar b + \bar t H^+ b$. Also $H^\pm$
production at hadron colliders can be studied through Drell-Yan processes for pair production, $qq \to H^+H^-$, the associated
 production with $W$ bosons, $qq \to H^\pm W^\pm$, and pair production through the gluon-gluon fusion
 $gg \to H^+ H^-$. 
The prominent decay modes for $H^\pm$ are $H^\pm \to tb$, $H^\pm \to \tau \nu$ and $H^\pm \to W^\pm h$. Here, with the
allowed vertices in the 2HDM, one could think of channels where $H^\pm$ couples with $H$ and further $H \to h \chi\chi$
gives opportunities to study this mode in terms of $\chi$ or missing energy aspects. The decay mode $H^\pm \to W^\pm H$
as discussed before can be included in the studies as a prominent channel.
The phenomenology of $H^\pm$ also depends on whether (i) $m_h < m_H < m_A$ or (ii) $m_h < m_A < m_H$, 
since $m_{H^{\pm}}$ could be consider as heavy as $m_A$.

As noticed, in all our studies we are considering an extra scalar $\chi$ as a dark matter candidate - as a signature of 
missing energy. But while considering various other excesses in data it may not be suitable to consider $\chi$ as a scalar.
So we need to characterise $\chi$ in other possible theories which could give appropriate reasons to consider the
scale factors $\beta_i$'s introduced in the previous discussion, since the production (decay) modes for $H/A$ are 
through $gg$ ($\gamma\gamma$) loops and there $\chi$ may be introduced in loops. 
Suppose the events with $\gamma\gamma$
excess scenarios, one may introduce an extra particle in the studies, with possible ranges of $m_A$, and then $A$ 
could be produced or decay via loops of massive coloured fermions. 
These extra particles could not be a fourth generation of the SM particles. 
Simple possibilities for these extra fermions maybe: (i) a single vector-like charges 2/3 quark.
(ii) an isospin doublet of vector-like charge 2/3 and -1/3 quarks, (iii) an isodoublet and two singlet charge 2/3 and -1/3
quarks, and (iv) a complete vector-like generation including leptons as well as quarks. In this respect we should 
consider all four characteristics of $\chi$, as vector-like fermions (VLF). Similar studies can follow for $W^\pm$ and
$Z$ decay modes of $A$.           

\subsection{A Two-Higgs doublet model}
\label{2hdm}
Now by knowing the requirements discussed above, we need to investigate a particular scenario of a 2HDM to 
accommodate and constrain the parameter space of this model. To start with the production mechanism of the
scalars $h$, $H$ and a pseudo scalar $A$, we should notice the fact that at the LHC the dominant mode of
production is through $gg$-fusion. However, $gg\phi$, ($\phi = h, H, A$) tree level couplings are not available in 2HDMs
was the case for the SM Higgs also, and is only possible through various quark-loop contributions. The generalised functional 
form is given as (see, e.g. Ref.~\cite{gunion})
\begin{align}
\tau \left[ 1 + \left( 1-\tau \right) f(\tau) \right] \to \tau \left[ \xi^\phi + \left( 1-\tau \xi^\phi \right) f(\tau) \right],
\label{func}
\end{align}  
where $\xi^\phi = 1$ for $\phi = h, H$ and $\xi^\phi = 0$ for $\phi = A$. $\tau$ is the ratio of masses of particles involved
in the one-loop calculation. A similar functional form is true for $\gamma \gamma \phi $ vertices with a multiplication by
a factor of $- 2$ in Eq.~(\ref{func}), with appropriate $f(\tau)$. A proper calculation demands involved mixing
parameters and couplings in that particular type of 2HDM model. 
So effectively depending on a choice of $\tau$ limits, $f(\tau)$ is correspondingly chosen and therefore we can 
incorporate those factors in a coupling parameter, $\kappa_{gg\phi}$ and $\kappa_{\gamma\gamma \phi}$.
Accordingly we can write the Lagrangian for these interactions as,
\begin{align}
{\cal L}_{gg\phi} =&\, -\frac{1}{4} \kappa_{ggh} G_{{\mu\nu}^b} G^{\mu\nu b} h
 - \frac{1}{4} \kappa_{ggH} G_{\mu\nu}^b G^{\mu\nu b} H
 - \frac{1}{4} \kappa_{ggA} G_{\mu\nu}^b {\tilde{G}^{\mu\nu b}} A, \\
{\cal L}_{\gamma \gamma \phi} =&\, -\frac{1}{4} \kappa_{\gamma\gamma h} F_{\mu\nu} F^{\mu\nu} h
 - \frac{1}{4} \kappa_{\gamma\gamma H} F_{\mu\nu} F^{\mu\nu} H
 - \frac{1}{4} \kappa_{\gamma\gamma A} F_{\mu\nu} {\tilde F}^{\mu\nu} A.
\end{align}
In our choice of a 2HDM, $\kappa_{gg\phi}$ and $\kappa_{\gamma \gamma \phi}$ are give as
\begin{align}
\kappa_{ggh} =& \frac{\alpha_s}{3\pi v} \,\frac{3}{4} \,{\cal F}_{12}\left[\left( \frac{m_h}{2m_t} \right)^2 \right], \,\,
\kappa_{\gamma\gamma h} = \frac{e^2}{4 \pi^2 v} \,\frac{47}{18} \,{\cal S}_{12}\left[\left( \frac{m_h}{2m_W} \right)^2,\left(\frac{m_h}{2m_t} \right)^2 \right], \\
\kappa_{ggH} =& \,\beta_g^H \frac{\alpha_s}{3\pi v},  \,\,
\kappa_{\gamma\gamma H} = \beta_\gamma^H \frac{e^2}{4 \pi^2 v}, \,\, \text{and} \,\,
\kappa_{ggA} = \beta_g^A \frac{\alpha_s}{3\pi v},  \,\,
\kappa_{\gamma\gamma A} = \beta_\gamma^A \frac{e^2}{4 \pi^2 v}.
\end{align}
Functions ${\cal F}_{12}$ and ${\cal S}_{12}$ are approximately Eq.~(\ref{func}) with appropriate $f(\tau)$ in 
a proper limit of $\tau$ (e.g. see~\cite{gunion}).
The multiplicative factors $\beta^{H,A}_{g(\gamma)}$ are introduced to control the rates for $gg$ and
$\gamma\gamma$-channels.  Interactions with electroweak vector bosons ($V$) $W^\pm, Z$ and the photon field, 
$A_\mu$, with $\phi$ and $H^\pm$, are given as
\begin{align}
{\cal L}_{VV\phi} =&\, \frac{2 M_W^2}{v} \cos(\beta - \alpha) W^+_\mu W^{- \mu} H
 + 2 \frac{M_W^2}{v} \left(\sin(\beta - \alpha) \right) W^+_\mu W^{- \mu} h \notag \\
& + \frac{M_Z^2}{v} \cos(\beta - \alpha) Z_\mu Z^\mu H
 + \frac{M_Z^2}{v} \left(\sin(\beta - \alpha)\right) Z_\mu Z^\mu h,
 \end{align}
 and,
 \begin{align}
{\cal L}_{V\phi\phi} =&\,  \frac{M_W}{v\,\cos\theta_W} \sin(\beta - \alpha) Z_\mu
\left(A\partial_\mu H - H\partial_\mu A \right)
 + \frac{M_W}{v\,\cos\theta_W} \cos(\beta - \alpha) Z_\mu (A\partial_\mu h - h\partial_\mu A) \notag \\
& + i \frac{M_W}{v} \frac{(2\,\cos^2\theta_W - 1)}{\cos\theta_W} Z_\mu \left(H^-\partial_\mu H^+  - H^+\partial_\mu H^- \right)
 + i e A_\mu \left(H^-\partial_\mu H^+  - H^+\partial_\mu H^- \right) \notag \\
& + \left[ i \frac{M_W}{v} \sin(\beta - \alpha) \left(W^{-\mu} H\partial_\mu H^+  - W^{-\mu} H^+ \partial_\mu H \right) 
+ \text{h.c} \right] \notag \\
& + \left[ i \frac{M_W}{v} \cos(\beta - \alpha) \left( W^{-\mu} h\partial_\mu H^+  - W^{-\mu} H^+ \partial_\mu h \right) 
+ \text{h.c} \right] \notag \\
& + \left[ \frac{M_W}{v} \left(W^{-\mu} A\partial_\mu H^+  - W^{-\mu} H^+ \partial_\mu A \right) + \text{h.c} \right].
\end{align}
We choose the most canonical 2HDM, that is the Type-II 2HDM for our analyses. In a Type-II
framework, the Yukawa terms look like
\begin{align}
{\cal L}^{Y}_{h} =&\,  -\frac{1}{v}\left[ \frac{\cos\alpha}{\sin\beta} \sum_{q_u} y_{m_{q_u}} q_u \bar q_u h 
+\frac{\sin\alpha}{\cos\beta} \sum_{q_d} y_{m_{q_d}} q_d \bar q_d h \right],\\
{\cal L}_{H}^{Y} =&\,  -\frac{1}{v}\left[\frac{\sin\alpha}{\sin\beta} \sum_{q_u} y_{m_{q_u}} q_u \bar q_u H
+ \frac{\cos\alpha}{\cos\beta} \sum_{q_d} y_{m_{q_d}} q_d \bar q_d H \right], \\
{\cal L}_{A}^{Y} =&\,  - \frac{i}{v} \left[ \cot\beta
  \sum_{q_u} y_{m_{q_u}} q_u \gamma_5 \bar q_u A + \tan\beta \sum_{q_d} y_{m_{q_d}} q_d \gamma_5 \bar q_d A \right],  \\
{\cal L}_{H^\pm}^{Y} =&\,  \frac{1}{2} \Big[ \left(-y_{ut} \cos\beta + y_{ub} \sin\beta\right) \left(\bar t b H^+ + \bar b t H^-\right) \notag \\
&\,\,\,\,+   \left(y_{ut} \cos\beta + y_{ub} \sin\beta\right) \left(\bar t \gamma_5 b H^+ - \bar b \gamma_5 t H^-\right)\Big],
\end{align}
with $y_{ut} = \sqrt{2} y_{m_t}/(v \sin\beta)$ and $y_{ub} = \sqrt{2} y_{m_b}/(v \cos\beta)$. 
The relevant trilinear scalar interactions are part of the Lagrangian ${\cal L}_{\phi\phi\phi}$, 
\begin{align}
{\cal L}_{\phi\phi\phi} =&\,  - v \lambda_{h H^+ H^-} h H^+ H^- - v \lambda_{h H^+ H^-} H H^+ H^- 
- \frac{1}{2} v \lambda_{H h h} H h^2,
\end{align}
where the couplings have the following expressions:
\begin{align}
\lambda_{h H^+ H^-} =&\, \frac{-1}{2 v^2 \sin(2\beta)} \Big[  m_h^2 \cos(\alpha-3 \beta) + 3 m_h^2 \cos(\alpha+\beta) 
-4 m_{H^\pm}^2 \sin(2\beta) \sin(\alpha-\beta) \notag \\
&\qquad \qquad\qquad  - 4 M^2 \cos(\alpha+\beta)   \Big], \\
\lambda_{H H^+ H^-} =&\, \frac{-1}{2 v^2 \sin(2\beta)} \Big[  m_H^2 \sin(\alpha-3 \beta) + 3 m_h^2 \sin(\alpha+\beta) 
+4 m_{H^\pm}^2 \sin(2\beta) \cos(\alpha-\beta) \notag \\
&\qquad \qquad\qquad - 4 M^2 \sin(\alpha+\beta)   \Big], \\
\lambda_{H h h} =& \frac{-1}{2 v^2 \sin(2\beta)} \Big[  (2 m_h^2 + m_H^2) \cos(\alpha-\beta)\sin(2 \alpha) \notag \\
&\qquad \qquad\qquad - M^2 \cos(\alpha-\beta) (3 \sin(2\alpha) - \sin(2\beta))  
   \Big].
\end{align}
Here $M^2$ is the shorthand notation for $m^2_{12}/(\sin\beta \cos\beta)$. 
This completes the whole discussion on the formalism 
of a specific 2HDM, which could further be used for most of the phenomenological studies to explain the experimental data; 
and vice-versa from experimental data the parameters of this model can be constrained and hence search studies will 
follow. 

In several places we discussed an additional particle $\chi$ which could be characterised 
as a DM candidate scalar or a scalar or maybe vector-like fermions. Lets consider the case when $\chi$
is a DM candidate, then the 2HDM model potential defined in Eq.~(\ref{pot2hdm}) needs 
an additional term. 
One can consider $\chi$ as a gauge-singlet scalar, as a DM candidate if its mixing with the doublets $\Phi_1$
and $\Phi_2$ can be prevented by the introduction of some discrete symmetry. One such symmetry is a $Z_2$
under which $\chi$ is odd and all other fields are even. This also ensures the stability of $\chi$.
Thus the most general potential consistent with the gauge and $Z_2$ symmetries can be written as:
\begin{align}
V\left(\Phi_1,\Phi_2,\chi\right) =  V\left(\Phi_1,\Phi_2\right) + \frac{1}{2}m^2_{\chi}\chi^2
 + \frac{\lambda_{\chi_{1}}}{2}\Phi_1^\dag \Phi_1 \chi^2 + 
 \frac{\lambda_{\chi_{2}}}{2}\Phi_2^\dag \Phi_2 \chi^2 + 
 \frac{\lambda_{\chi_{3}}}{4}(\Phi_1^\dag \Phi_2 + \text{h.c}) \chi^2.
\end{align}
And hence the complete interaction terms with $h$ and $H$ will follow as
\begin{align}
{\cal L}_{\phi\chi\chi} =  - \frac{1}{2} v \lambda_{h \chi \chi} h \chi^2 - \frac{1}{2} v \lambda_{H \chi \chi} H \chi^2,
\end{align}
where the couplings are given as,
\begin{align}
\lambda_{h \chi \chi} =& \lambda_{\chi_{1}} \cos\beta \sin\alpha - \lambda_{\chi_{2}} \sin\beta \cos\alpha - \frac{1}{2}\lambda_{\chi_{3}}\cos(\beta + \alpha), \\
\lambda_{H \chi \chi} =& -\lambda_{\chi_{1}} \cos\beta \cos\alpha - \lambda_{\chi_{2}} \sin\beta \sin\alpha - \frac{1}{2}\lambda_{\chi_{3}}\sin(\beta + \alpha).
\end{align} 
Similarly, we follow the studies for other characteristics of $\chi$ based on gauge and symmetry issues. 
For most of the details on 2HDMs we refer to Refs.~\cite{gunion,Branco:2011iw}. 
        
\section{Summary}
\label{conc}
We briefly reviewed Higgs boson physics with an extension of an additional Higgs-boson-doublet in the SM.  
Also we talked about a study where 
two hypothetical scalars $H$ and $\chi$ with $2 m_h < m_H < 2 m_t$ and $m_\chi \approx m_h/2$,
were introduced to study the $p_T$ spectrum of $h$, via a process $pp \to H \to h \chi \chi$.
The required vertices are defined in an effective way where $H \to h\chi\chi$ branching fraction may be assumed to be large
or small depending on the choice of values of the coupling $Hh\chi\chi$.
The introduction of a new scalar, $S$, has been discussed to explain large branching ratios for $H \to h\chi\chi$
decays with an effective  $hHS$ vertex through a $S\to \chi\chi$ decay mode, with 
$m_h \lesssim m_S \lesssim m_H-m_h$ and $m_S > 2 m_\chi$. 
The introduction of a $HSS$ vertex leads to the $H \to SS$ decay.

We discussed salient features of phenomenological aspects of CP-odd scalar, $A$, in the 2HDMs. 
Various channels for phenomenological interests in multi-lepton final sates with same and opposite-sign 
lepton selection with $Z$+jets+MET, have been discussed.  
In this respect we discussed how the associated parameters with the charged Higgs $H^\pm$ under prominent 
decay modes to $W^\pm H, H \to h \chi \chi$ and $t\bar b$ will be probed at the LHC. 

It was discussed that $\chi$ need not always be taken as a DM candidate, and its characteristic for different 
phenomenology may need to be characterised as a scalar or as vector-like fermions.

A formalism was presented in a particular type of 2HDM. An appropriate Lagrangian for this
model was presented in detail, which further used various studies, as discussed, for the phenomenology of extra
scalars. Considering $\chi$ to be a DM candidate, interactions terms in a specific 2HDM scenario is
presented. 

\section*{Acknowledgements}
 T.M. is supported by funding from the Carl Trygger Foundation under
contract CTS-14:206 and the Swedish Research Council under contract
621-2011-5107.

\end{document}